# Coexistence of exchange bias training effect and spin-orbit torque in IrMn-layer/ferromagnetic-ribbon heterostructures via magnetoimpedance effect

M. R. Hajiali[1, †, *], L. Jamilpanah[1, †], J. Shoa e Gharehbagh[1], S. Azizmohseni[1],
A. Olyaei[2], G. R. Jafari[1], M.M. Tehranchi[1,2] and Majid Mohseni[1, *]

[1]*Faculty of Physics, Shahid Beheshti University, Evin, 19839 Tehran, Iran*
[2]*Laser and Plasma Research Institute, Shahid Beheshti University, Evin, 19839 Tehran, Iran*

We investigate a possible correlation between spin orbit torque (SOT) and exchange-bias (EB) in IrMn-layer/ferromagnetic-ribbon heterostructure by performing magnetoimpedance (MI) measurements. To uncover this correlation, we benefit from EB training effect probed by MI effect at room temperature. A damping-like SOT driven by ac current through the antiferromagnetic IrMn applies to the ferromagnetic ribbon layer, determined by MI magnetic field and frequency sweeps. Importantly, magnitude of the SOT is observed to remain intact against EB training and decrease of EB through alternative magnetic field sweep cycles. Our results pave the way to better elucidate the EB effect, EB training and the SOT, useful for future spintronic elements.

**Keywords:** Spintronics, *Exchange bias, Training effect, Spin-orbit torque*, *Magnetoimpedance effect, Skin depth*

[*]Corresponding author's email address: mrh.hajiali67@gmail.com, m-mohseni@sbu.ac.ir
[†] These authors contributed equally.

# I. INTRODUCTION

Recently antiferromagnetic (AFM) materials have attracted great attention in the field of spintronics, due to their attractive features such as insensitivity to external magnetic fields, ultrafast dynamics and generation of large magneto-transport effects[1–3]. Presence of prominent magneto-transport effects in devices composed of AFM thin films has coordinated significant impacts in devices proposed for giant and anisotropic magnetoresistance[4,5], spin-orbit torque (SOT)[6], direct and inverse spin Hall effect (SHE)[7,8], and current-induced domain wall motion[9]. Of the various AFM materials, noncollinear chiral ones have a special place, due to their remarkable structural, magnetic, and electrotransport properties. A triangular spin configuration of AFM compounds gives rise to a large anomalous Hall effect (AHE)[10], magneto-optical Kerr effect[11,12] and SHE[13]. Both the AHE and SHE are favorite candidates to drive the magnetization dynamics of a ferromagnet in which spin currents can be used to manipulate magnetic moments. Because AFMs can have a large conventional SHE, they can thus exert a large magnitude SOT on an adjacent ferromagnet (FM), that was confirmed by several experiments[14–16].

On the other hand, the exchange-bias (EB) effect, known historically as a shift in the hysteresis loop ($H_{EB}$) of a ferromagnet arising from interfacial exchange coupling between adjacent FM and AFM layers[17–19], is an integral part of spintronic devices. This interfacial coupling has been studied intensively in the past two decades because of its applications in magnetic devices such as spin valve and magnetic random access memory (MRAM). Beside the hysteresis loop shift and coercivity enhancement, the EB phenomenon also exhibits asymmetry in the magnetization reversal process and training effect (TE)[20–25]. The TE refers to the gradual and monotonous degradation of both $H_{EB}$ and coercivity ($H_C$) to equilibrium values during consecutive hysteresis loop measurements after field growth or field cooling. It is generally accepted that, the TE arises due to irreversible changes in the magnetic microstructure of the AFM layer, as its spin texture re-arranges with each magnetization reversal of the FM layer.

Therefore, in an FM/AFM structure both SOT and EB effects coexist and affect the magneto transport. However, the contribution of interfacial exchange coupling to the amplitude of the SOT in the FM/AFM interface remains controversial. From this point of view, Tshitoyan et al.[26] have demonstrated that there is a direct link between the magnitude of the EB and spin torque efficiency whereas Saglam et al.[27] have seen that the SOT is independent from the EB direction in a NiFe/IrMn bilayer. In addition, Zhang et al.[13] found that there is no correlation between the in-plane EB and the magnitude of spin Hall angle. Therefore, to reconcile this issue, finding a correlation between the EB strength and SOT is yet under debate as one of the most important aspects in AFM spintronics.

In this work, we try to uncover presence of such a correlation using the MI effect in high permeability FM/AFM bilayer. The studied heterostructure is made of an amorphous FM $Co_{68.15}Fe_{4.35}Si_{12.5}B_{15}$ ribbon and a thin AFM layer of polycrystalline $Ir_{20}Mn_{80}$ (IrMn). In the MI effect, changes in the electrical impedance of a conducting FM with high transverse magnetic permeability ($\mu_t$) in the presence of a static magnetic field can be monitored[28–32]. By applying an external magnetic field, the skin depth ($\delta = (\rho/\pi\mu_t f)^{1/2}$) of applied ac current changes due to a change in the $\mu_t$ that alters the impedance. Accordingly, the impedance of the FM ribbon is a function of frequency, driving current, and the external dc magnetic field ($H$) through $\mu_t$ and $\delta$. Very recently we proposed that impedance spectroscopy can be used for detection of the SOT resulting from the SHE in a Pt-layer/magnetic-ribbon heterostructures[33]. Furthermore, we showed that presence of the damping-like (DL) torque arising from the Pt layer deposited on the FM not only changes the MI response, but also tends to vary the transverse anisotropy of the magnetization, that was evidenced by a different impedance frequency shifts.

In this work, we observe the DL torque originating from an AFM IrMn deposited on magnetic ribbon due to the conventional SHE. In addition, in order to investigate the correlation between the magnitude of the EB and SOT, we study details of the EB and TE in our system. In fact, the MI effect probes the spin texture at the skin depth profile which is approximately less than 100 nm and is close to the interface. Interestingly, the MI is used to observe both the EB and TE, and also is able to probe the SOT. Hence, by using the MI spectroscopy, it is expected to find a correlation between the EB and SOT. Our results can be used for development of simple methods for study of fundamental and experimental spintronic phenomena.

## II. EXPERIMENTAL DETAILS

Amorphous Co-based FM ribbons (1 mm width, 40 mm length and ~20 µm thickness) was prepared by a conventional melt-spinning technique. Before deposition of the IrMn layer, about 40 nm of the ribbons surface was etched via Ar plasma to have a clean and oxygen free surface. Immediately after that, the IrMn thin layer with thickness of 20 nm was deposited on the soft surface (shiny side) of those ribbons in the presence of the Ar with a pressure of 5 mTorr, base pressure better than $5\times10^{-6}$ Torr and growth rate of 3 nm/minute. The IrMn-layer/ferromagnetic-ribbon heterostructures are continuous and do not have grain chains like electrodeposited nanowires[34,35]. In order to induce the EB, the as-deposited sample was subsequently annealed at 280 °C for 1 h under vacuum conditions ($4\times10^{-3}$ Torr) in the presence of in-plane magnetic field and then cooled down to room temperature. The applied magnetic field during the annealing and the cooling process was set to 230 Oe and applied in the longitudinal direction of the sample. Magnetic hysteresis loops were measured using longitudinal magneto-optic Kerr effect (MOKE). The magnetic structure of the IrMn layer was characterized by the Grazing incident x-ray diffraction (GID) method. For the MI measurements a 4 cm ribbon with preserved width and length was used and an external magnetic field produced by a solenoid applied along the ribbon axis and the impedance was measured by means of the four-point probe method. The ac current passed through the longitudinal direction of the ribbon with different frequencies supplied by a function generator (GPS-2125), with a 50 Ω resistor in the circuit. The impedance was evaluated by measuring the voltage and current across the sample using a digital oscilloscope (GPS-1102B). The MI ratio is defined as $MI\% = \frac{Z(H) - Z(H_{max})}{Z(H_{max})} \times 100$; where $Z$ refers to the impedance as a function of the external field $H$. The $H_{max}$ is the maximum field applied to the samples during the MI measurement. All measurements were carried out at room temperature.

## III. RESULTS AND DISCUSSION
### A. MOKE and GID measurements

In order to examine the presence of the EB effect and occurrence of the TE in our sample, we first measured magneto-optical Kerr effect (MOKE) hysteresis loop of FM/IrMn heterostructures. Figure 1(a) illustrates the consecutive MOKE hysteresis loops (with magnetic field applied along the induced EB direction) with n = 1, 2, 3 and 4 (n= number of field sweep), for the magnetically annealed (280 °C /230 Oe) FM ribbon/IrMn sample. Inset of the figure shows the hysteresis loop for the FM ribbon that does not show any shift. As shown, the hysteresis loop for n=1 clearly exhibits the EB behavior as a shift of the loop towards negative values, by $|H_{EB}|$ =2.4 Oe. During the consecutive M–H loop measurements, the TE was observed for the annealed IrMn/ ribbon sample. One can note that the substantial decrease in $H_{EB}$ takes place only between the first two consecutive hysteresis loops and also the shift is more prominent in the descending branch because in our measurement the loop starts at positive saturation.

In order to determine of the magnetic structure of the IrMn layer, we performed GID measurement. Figure 1(b) shows the GID profile of the ribbon, ribbon/IrMn and the magnetically annealed (280 °C /230 Oe) ribbon/IrMn heterostructure. The calculated powder diffraction pattern of IrMn$_3$ was also attached at the top of the Fig. 1(b). GID of the ribbon/IrMn before field annealing shows that the IrMn is crystalized in (111) crystalline direction of IrMn$_3$. After field annealing, the peak related to the (220) crystalline direction of IrMn$_3$ does appear too. Both observed peaks are related to the crystallographic γ-phase.

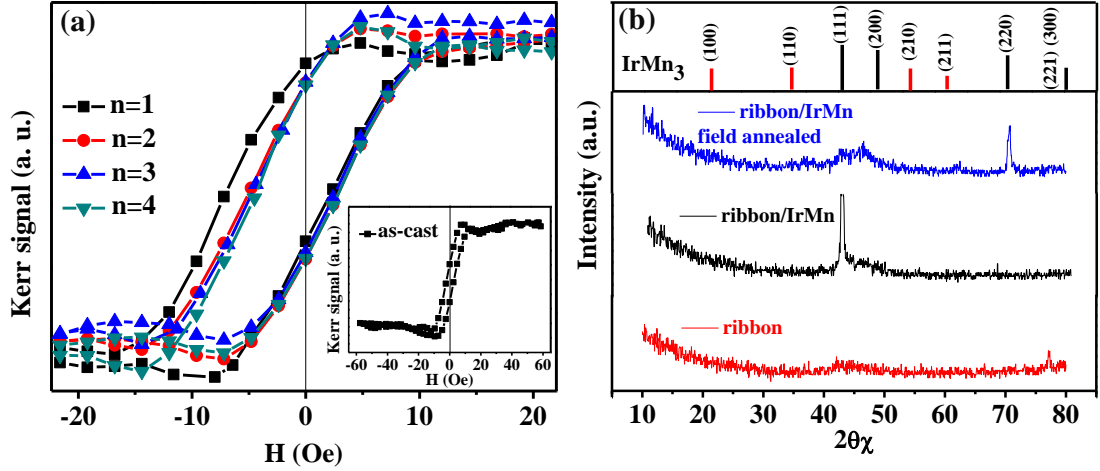

FIG. 1. (a) Successive MOKE hysteresis loops of the magnetically annealed (280 °C /230 Oe) FM ribbon/IrMn (20 nm) heterostructure at room temperature. Inset displays the hysteresis loops for the FM ribbon that does not show any shift so that, for field annealed ribbon/IrMn heterostructures, the hysteresis loop clearly exhibits the EB effect by $|H_{EB}|$ =2.4 Oe. Also the hysteresis loop for n=1 clearly exhibits the TE between the first two consecutive hysteresis loops. (b) Grazing incident x-ray diffraction profiles of the ribbon, ribbon/IrMn and the magnetically annealed (280 °C /230 Oe) FM ribbon/IrMn (20 nm) heterostructure. The calculated powder diffraction pattern of IrMn$_3$ was also attached at the top of the figure.

Formation of polycrystalline (111) textured IrMn$_3$ is expected similar to results observed by others when they sputtered IrMn on amorphous SiO$_2$[13]. Because, the ribbon has amorphous structure too. Identification of the γ-phase and the Cu3Au-type phase (L12) in AFM/FM bilayers deposited by magnetron sputtering is possible via GID. In the works of Imakita et al.[36,37] and Tsunoda et al.[38], the authors verified the presence of the ordered L12-IrMn by evaluating the peak intensities of the (110) and (211) superlattice reflections in their measurement. Structurally the samples studied in our work are similar to the ones studied by these authors. In Fig. 1(b) the black bars correspond to those reflections which are present in both γ-phase and ordered L12-IrMn, and the indicated red bars correspond to the additional superlattice diffraction peaks resulting from the L12-IrMn phase. Because there is no (110) and (211) observed peaks, which correspond to superlattice reflections of the L12-IrMn phase, indicates that our crystallized IrMn has a γ-phase structure.

### B. Impedance measurements with magnetic field sweep

In previous section we showed that the EB is present in our samples. Here, in order to elucidate the effect of the EB coupling on the MI effect, we carry out field sweep impedance measurements. By applying an ac charge current with frequency $f$=10 MHz to the samples, we investigate how the impedance of the heterostructures changes as a function of the external magnetic field. So far, effect of the EB coupling strength on the MI was investigated theoretically and experimentally in different structures e.g. in thick FM ribbon/oxide layer[39] and also NiFe/FeMn[40] and NiFe/IrMn[41,42] thin film multilayers. Due to presence of the EB in those systems, shifts were observed in both the hysteresis loop and the MI response, and the MI response showed an *asymmetric magnetoimpedance* (AMI). Generally, the MI as a function of the field has shown asymmetric response against field polarities by applying additional bias field[43] or current[44], and EB [39]. For thick FM ribbon/oxide layer, the exchange interaction between the amorphous bulk and the surface crystalline regions leads to an effective unidirectional anisotropy which is responsible for the AMI. For thin multilayers, the MI with significant EB effect was seen in high frequency regime because of low permeability of these layers in low frequency (MHz). However, there is no report on the presence of the EB in thick FM ribbon/AFM probed via the MI effect within the MHz frequency ranges.

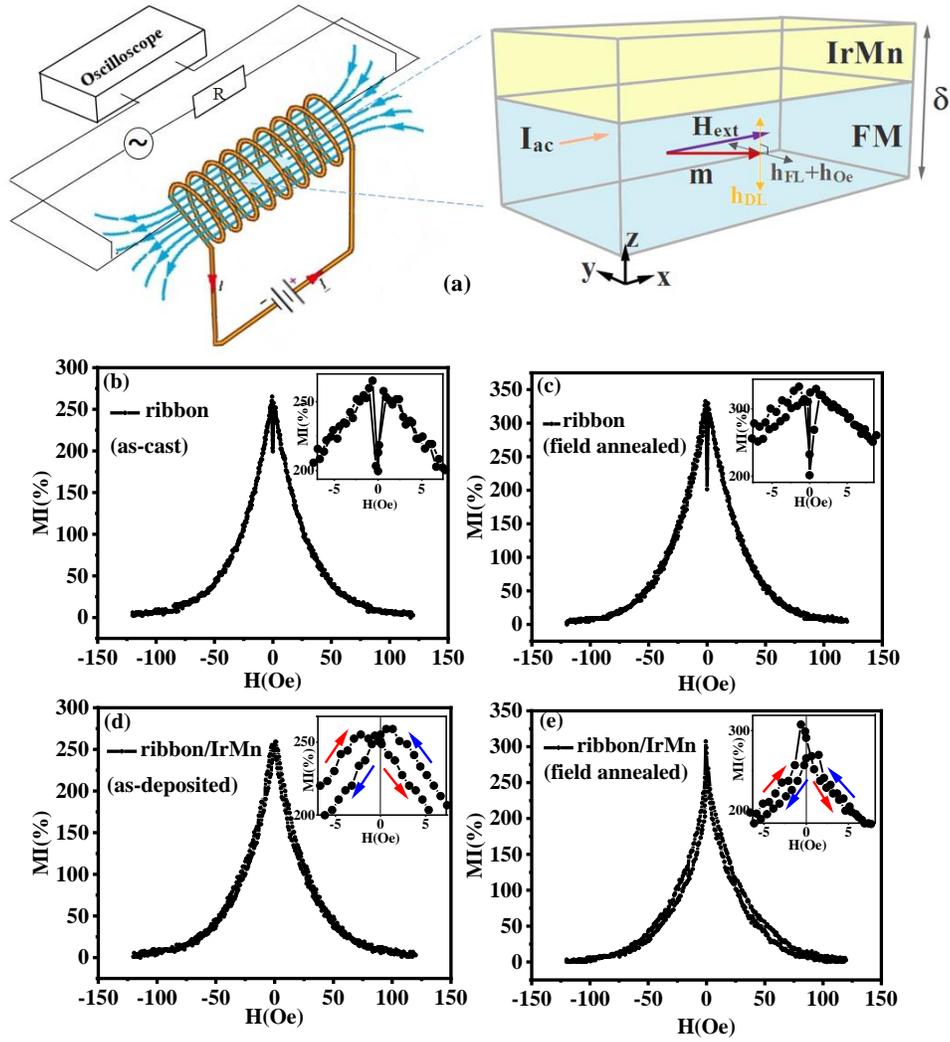

FIG. 2. (a) Schematic illustration of the measurement setup for the MI response and the structural conditions of a ribbon/IrMn heterostructure. We should note here that this scenario happens at the skin depth $\delta$ of the ribbon. (b-e) Comparison of the MI response of the ribbon and the ribbon/IrMn samples at frequency of $f$ = 10 MHz and an applied current of $I$= 66 mA. Inset shows the MI response around zero magnetic fields.

Fig. 2 (a) show schematic illustration of the measurement setup for the MI response and the structural conditions of a ribbon/IrMn heterostructure. Fig. 2(b-c) and (d-e) show field sweep impedance measurements at frequency of $f$ = 10 MHz for ribbon (as-cast and field annealed) and ribbon/IrMn heterostructures (as-deposited and field annealed) samples, respectively, with a current of $I$ = 66 mA applied to the samples. Three main differences between the as-cast ribbon and the IrMn-deposited ribbons can be seen in MI plots.

- *i)* MI% decreases from 265% for the as-cast ribbon to 259% for as-deposited ribbon/IrMn, and from 325% for the field annealed as-cast ribbon to 314% for the field annealed ribbon/IrMn samples. The decrease of MI response for IrMn-deposited samples is associated with a decrease of the transverse magnetic permeability, $\mu_t$. Due to presence of the EB effect in the IrMn/ribbon interface, which pins spins of the ribbon with additional anisotropy and reduces the $\mu_t$ and thereby results in a decrease in MI ratio. Moreover, it can be seen in Fig. 2(b) and (d) that the as-cast ribbon exhibits a double peak behavior whereas the IrMn deposited ribbon shows a single peak one. This trend is also repeated for field annealed samples in which their MI results are displayed in Fig. 2(c) and (e). In general, the observed single or double-peak behavior is associated with the longitudinal or transverse magnetic anisotropy with respect to the external field direction[45,46]. Thus, we conclude that the IrMn layer changes the transverse magnetic anisotropy of ribbon.
- *ii)* we see a hysteretic MI behavior for IrMn deposited samples (Fig. 2(d-e)), contrary to that as-cast ribbons, potentially implying presence of an antiferromagnetic coupling of the bias field with magnetization inside the amorphous phase[47]. Furthermore, in our previous work[48], it was shown that the hysteretic behavior of the MI response of the ribbon depends on strength of magnetic exchange coupling in the bilayer. Accordingly, here in our sample (IrMn/ribbon), exchange coupling at the interface between IrMn and ribbon can cause the same effect and in turn the hysteretic behavior of MI response can be observed.
- *iii)* In ribbons coated with IrMn, by sweeping the magnetic field from positive to negative values (inset of Fig. 2(e)), we observe an asymmetric single peak of the MI response with a nonzero $H$ (2.4 Oe). The amount of this shift is similarly seen for magnetization loop as an EB shift ($|H_{EB}|$ =2.4 Oe). It should be noted that the shift of MI curve occurs over the whole range of frequencies (1-20 MHz, not shown here). As mentioned before, asymmetric MI (AMI) response due to present of the EB[39] is interpreted as the exchange interaction between the amorphous bulk and the surface crystalline regions, forming an effective unidirectional anisotropy. Here, as represented in Fig. 2 (e), the exchange interaction between the IrMn and the surface of ribbon can lead to an effective unidirectional anisotropy which is responsible for the AMI.

## C. Training effect probed by MI

In section III.A, using the MOKE measurement, we showed that the TE occurs for the field annealed ribbon/IrMn samples. In this section, the TE is evaluated through the MI effect. Toward this, we have further examined the successive field cycling while performing MI measurements for the same sample. Fig. 3(a-d) shows consecutive MI curves for mentioned samples at frequency of $f$ = 10 MHz, corresponding to first to fourth sequential MI measurements labeled by n = 1 to n = 4, respectively. We can observe that by repeating the MI measurements continuously, the peaks which appear at positive and negative low magnetic fields move towards $H$= 0 Oe. In addition, the asymmetric behavior of the MI peaks disappears. This is obviously a consequence of the TE at the interface of the ribbon/IrMn. When the MI measurement is repeated continuously, the strength of exchange interaction decreases due to TE and finally MI effect shows symmetric behavior. So the rearrangement of the magnetic moments of the AFM at the interface plays the main role here. By further repeating the measurement, no additional changes appear in the MI plot.

It is worth mentioning that by repeating the measurement from the first to fourth cycle, the MI% increases as well. As mentioned before, the MI% is dominated by the strength of $\mu_t$. By further repeating the MI measurement and rearrangement of the magnetic moments based on the TE as the EB vanishes, the $\mu_t$ of the ribbon increases due to unpinning and consequently the MI% increases. Therefore, we can observe the TE evidenced by a peak shift and an increase in the MI% (here it is (52%)). We repeat the experiments for a field annealed ribbon without IrMn and we observe no variation in MI response, even by repetitive measurements.

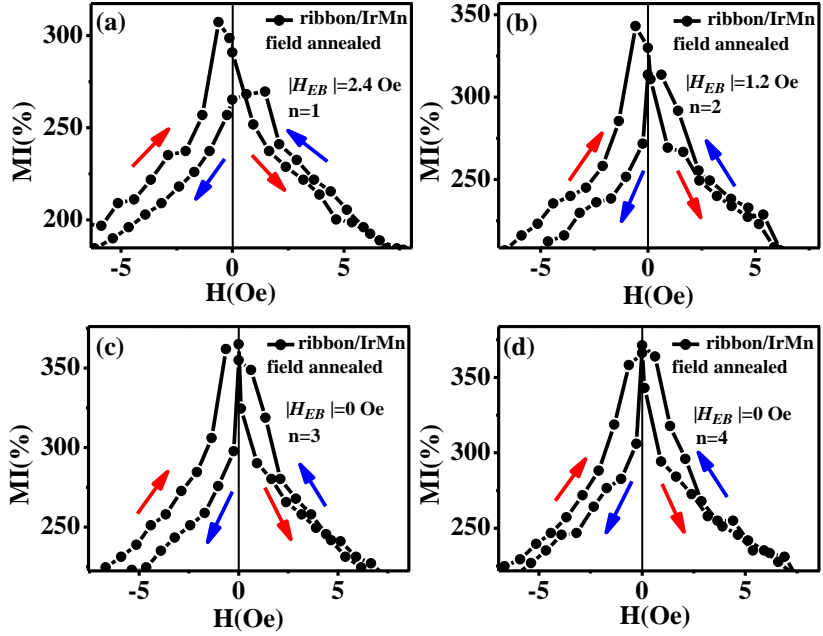

FIG. 3. MI response of field annealed ribbon/IrMn for 4 consecutive MI loops at $f$ =10 MHz. By repeating the MI measurements continuously (n = 1 to n = 4), the peak which is appearing at low magnetic fields moves towards the $H$= 0 Oe and finally the MI response at n=4 shows increased value with a symmetric and single peak behavior.

### D. Detection of SOT driven magnetoimpedance and independence of SOT from EB

In this section, we report frequency sweep MI measurements carried out right after each cycle of MI measurement under magnetic field sweep. These measurements reveal two conclusions which we will discuss separately in the following.

*Effect of SOT on MI due to SHE:* The results for frequency sweep MI measurements after three field sweeps are shown in Fig. 4(a-c), for as-cast ribbon, ribbon/IrMn and field annealed ribbon/IrMn samples (in the presence of an external field of 120 Oe). Here, the ac currents with different peak to peak amplitudes of $I$ = 33, 66 and 99 mA applied. It is noted that for all investigated samples, with increasing frequency, the maximum MI ratio first increases, reaches a maximum at a particular frequency, and then decreases in higher frequency ranges. The frequency dependent MI effect is explained considering a relative contribution of DW motion and moment rotation to the $\mu_t$[49,50]. As the frequency increases, the contribution of DW motion is damped due to presence of the eddy current and moment rotation becomes dominant. Also, this behavior is interpreted by the change in the internal part of the ribbon inductance due to change in permeability and eddy currents at different frequencies[49]. Therefore, a peak in the impedance of the ribbon versus frequency can be observed.

As can be seen in Fig. 4, the impedance peak frequency is shifted towards higher values, for either of IrMn coated samples with respect to the impedance of as-cast ribbon. The peak frequency under applied ac current of 99 mA is 7 MHz for as-cast ribbon and shifts to 8.2 and 11 MHz for as deposited and field annealed samples, respectively. By increasing the amplitude of the ac current, we observe a systematic shift in the frequency of the impedance peak, as shown in Fig. 4(d). This shift is higher for the sample after annealing and no such shifts observed for the as-cast ribbon. This directly confirms the effect of SOT and its role in the impedance which we will discuss in the following.

As shown schematically in Fig. 2(a), in the presence of the IrMn layer, the oscillating electric current generates an oscillating spin current. This spin current flows into the adjacent FM layer and exerts two different types of oscillating SOTs; field-like (FL) torque $\mathbf{T}_{FL} \sim \mathbf{m} \times \mathbf{y}$ that is equivalent to an in-plane field $\mathbf{h}_{FL} \sim \mathbf{y}$ and DL torque $\mathbf{T}_{DL} \sim \mathbf{m} \times (\mathbf{y} \times \mathbf{m})$ that establishes an out-of-plane field $\mathbf{h}_{DL} \sim \mathbf{m} \times \mathbf{y}$, where $\mathbf{m}$ is the magnetization unit vector and $\mathbf{y}$ is the in-plane axis perpendicular to the current flow direction $\mathbf{x}$. The generated spin current and spin accumulation at the interface can result various effects such as conventional SHE[51] and Rashba-Edelstein effect (REE). Generally, the REE and SHE are known as two origins of the $T_{FL}$ and $T_{DL}$. However, the $T_{FL}$ due to the REE revealed in ultra-thin FM layer adjacent to heavy metals layer[52], and the $T_{FL}$ due to the SHE is very weak in metallic systems[53].

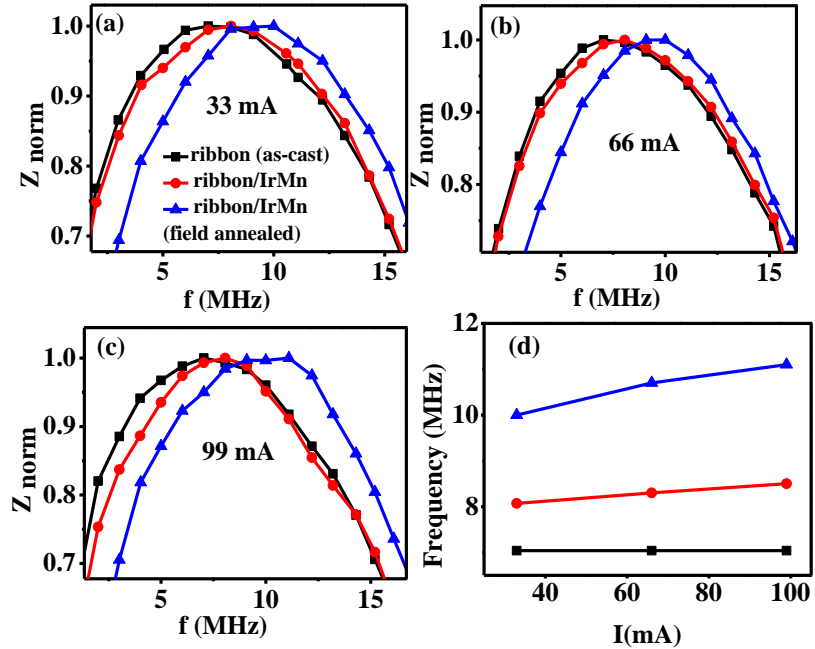

FIG. 4. (a-c) Frequency sweep impedance measurement of the ribbon, ribbon/IrMn and the field annealed ribbon/IrMn in the presence of an external field of 120 Oe. The ac current amplitudes (peak to peak) are (a) $I = 33$ mA, (b) $I = 66$ mA and (c) $I = 99$ mA to observe a higher frequency shift for higher driving currents. (d) Peak frequency obtained from (a)–(c) versus ac current amplitude, indicating the higher slope for the sample after annealing and no such shifts observed for the as-cast ribbon.

The $T_{FL}$ duo to SHE and REE is negligible in our studied structure because of the metallic nature of ribbon/IrMn and large thickness of the FM ribbon. Now on, we consider that $T_{DL}$ comes primarily from the SHE in IrMn and $T_{FL}$ comes from the Oersted torque ($T_{Oe}$) due to the Oersted field. $T_{Oe}$ is generated from applied ac charge current that depends on the conductivity of each layer and skin depth $\delta$.

The frequency shift in MI is more prominent in field annealed samples that can be explained based on two main reasons. One is the change in the IrMn resistivity and the other is the SOT efficiency. To see the effects of annealing on the electrical properties of the sample, we measured the resistance of our samples before and after annealing and no changes were observed. Hence, the only parameter is the SOT efficiency which is directly related to interface spin transparency and AFM domain structure. Two things can be resulted from thermal annealing: first simultaneous changes in the bulk and/or interfacial magnetic orders in AFM, second the degree of crystallization, as well as interfacial mixing. These all influence the measured SOT efficiency. Generally, two phenomena are considered in limiting the interfacial spin transparency: (i) spin backflow (SBF) and (ii) spin memory loss (SML) due to spin scattering at the interface. By thermal annealing can observe strong variations in interfacial spin transparency due to enhanced SBF and SML by interfacial intermixing[54,55]. However, a very recent work[54] shows that SML at HM/FM interfaces increases significantly with interfacial spin orbit coupling (ISOC). Therefore, the variation of the SOTs with annealing is a direct consequence of degradation of the spin transparency of the interface by the ISOC that becomes stronger with annealing. This effect will cause a decrease in SOT and thus less frequency shift in annealed samples in our case. However, presence of such a large frequency shift in annealed samples suggests another mechanism for increasing SOT efficiency.

In practice, the strength of SOT of IrMn substantially decreased due to the averaging effect of randomly oriented crystallites and domains[7]. According to the GID result (Fig. 1b), crystallographic structure of IrMn is changed after field annealing and also field annealing causes the magnetic structure (AFM domain) of sample to change. In Ref. 13, the effects of field annealing in (001)-oriented antiferromagnetic $IrMn_3$ thin films, coupled to ferromagnetic NiFe layers is investigated. The underlying mechanism is reported to be based on the SHE and perpendicular field annealing leads to increase in the SOT when the domains are oriented. In our sample, when field annealing is carried out, applying external magnetic field orients the direction of the magnetization of the ribbon. Hence, due to presence of the exchange interaction across the AFM/FM interface, the magnetic configuration of the AFM will change[13,36]. Since the FM layer is magnetized to a unique direction, exchange coupling at the interface of FM and AFM layer forces the IrMn domains to follow a unique direction. The AFM domains were randomly oriented before annealing while after that, more aligned AFM domains are present[13,36]. Before annealing, as the domain configuration of the sample tends to be random, the resulting SOT will have lower impact. Consequently, after annealing, the spin polarization of the spin current from the SHE has a preferential direction due to unique domain orientation in IrMn. Therefore, this spin current from SHE will exert a significant SOT on the FM layer, results in a high frequency shift in MI.

Another possible scenario for increasing SOT efficiency in FM/IrMn heterostructure can be related to the presence of a magnetic spin Hall effect (MSHE), that has been proposed theoretically and experimentally in noncollinear $IrMn_3$ AFM[56,57] and $Mn_3Sn$[58]. The MSHE is a transverse spin current, odd under reversal of magnetic moments, and has a distinct symmetry and origin from the conventional SHE. GID result (Fig. 1b) shows that the polycrystalline $Ir_{20}Mn_{80}$ (IrMn) is formed by grains of $IrMn_3$[59,60], a noncollinear AFM that recently predicted to show the MSHE[56,57]. It is important to note that, generated spin current by MSHE is a consequence of a symmetry breaking caused by noncollinear magnetic structure of $IrMn_3$ and is odd under time reversal. Therefore, spin current from MSHE will exert a significant SOT on the FM layer. Although, we have no precise characterization for the presence of MSHE in our samples. Further experiments are needed to reveal the microscopic mechanism of the phenomenon.

## IV. Conclusion

In summary we probed the EB effect, TE and presence of the SOT through the MI effect in ribbon/IrMn heterostructure. Using MI measurements with consecutive field sweep repetitions, we observed the TE that reduced the EB field. The SOT originating from the AFM IrMn in the mentioned heterostructure was observed as a shift in peak frequency of the MI effect, even by removing the EB field using the TE. Importantly, magnitude of the SOT is observed to remain intact against EB training and decrease of EB through alternative magnetic field sweep cycles. Our results can be used for development of simple methods to study the fundamental magnetic effects for application in spintronic technologies.

Support from Iran National Science Foundation (INSF) is acknowledged.